# Isolated zero-energy flat-bands and intrinsic magnetism in carbon monolayers

Chaoyu He,[1,2,*] Shifang Li,[1] Yuwen Zhang[1], Zhentao Fu[2], Jin Li[1] and Jianxin Zhong[2,†]

[1] School of Physics and Optoelectronics, Xiangtan University, Xiangtan 411105, China

[2] Center for Quantum Science and Technology, Shanghai University, Shanghai 200444, China

Email of Corresponding Author: hechaoyu@xtu.edu.cn; jxzhong@shu.edu.cn

Flat-band in twisted graphene bilayer has garnered widespread attention, and whether flat-bands can be realized in carbon monolayer is an interesting topic worth exploring in condensed matter physics. In this work, we demonstrate that, based on the theory of compact localized states, a series of two-dimensional carbon allotropes with flat-bands can be achieved. Two of them named as 191-8-66-C-r567x-1 and 191-10-90-C-r567x-1 are confirmed to be dynamically stable carbon phases with isolated or weakly overlapped flat-bands at the Fermi-level. The maximum Fermi velocities of the flat-band electrons are evaluated to be $1\times10^4$ m/s and $0.786\times10^4$ m/s, both of which are lower than the Fermi velocity of the flat-band electrons in magic-angle graphene ($4\times10^4$ m/s). Furthermore, 191-8-66-C-r567x-1 has been confirmed to be a flat-band related magnetic half-metal with a magnetic moment of 1.854 μB per cell, while 191-10-90-C-r567x-1 is a flat-band related magnetic normal metal with a magnetic moment of 1.663 μB per cell. These results not only show that flat-bands can be constructed in carbon monolayer, but also indicate the potential for achieving metal-free magnetic materials with light elements based on flat-band theory.

## 1. Introduction

Crystal structure is one of the most fundamental and important concepts in condensed matter physics and materials science, as it largely determines the physical and chemical properties of condensed matter. It can be further simplified into lattice models under the tight-binding (TB) approximation to quickly calculate the dispersion relation for understanding the electronic property of materials. Two extreme cases classified according to effective mass, namely, the linear-dispersion band [1] holding massless Dirac fermion and the dispersionless flat-band [2-4] with infinitely heavy fermion have attracted wide research interests. The Dirac fermions have been experimentally detected in graphene monolayer [1] and they have been extensively studied. Although the flat-band has also been experimentally realized in magic-angle graphene [5, 6], only a small number of lattices confirmed to exhibit flat-bands [2-4, 7-9]. Flat-bands in real materials are still very rare [2-4, 10-13] for realizing exotic quantum phases, such as high-temperature fractional quantum Hall effects [14], Bose-Einstein condensation [15], Wigner crystallization [16, 17], high-temperature superconductivity [18] and ferromagnetism [12, 19]. That is to say, more flat-band lattices [20] are still expected to guide us in designing new flat-band systems, including electronic materials [4, 21], photonic crystals [22], and acoustic superstructures [23].

In electronic materials, isolated flat-bands near the Fermi-level are more interesting and those far away from the Fermi-level are less valuable. However, most flat-bands in previously proposed lattices are not situated at the Fermi-level and are often not isolated. It is challenging to isolate these flat-bands and engineer them to the Fermi level. In real crystalline materials, metal-organic frameworks (MOF), carbon-based frameworks (COF) and hydrogen-bond organic frameworks (HOFs) are excellent choices [24-28] for designing ideal flat-bands. For example, the lieb-like $sp^2$C-COFs and $sp^2$N-COFs [11, 12], the kagome-like cyclic graphene, cyclic graphyne and cyclic graphdiyne [17, 29], as well as the triphenyl-manganese (TMn) lattice [30] and $X_3(HITP)_2$ monolayer [31], are reported holding flat-bands with good flatness. Isolated flat-bands have also been realized in realistic materials of 2D hydrocarbons [13], holey graphene [32] and carbon nitride frameworks [33-35]. However, these flat-bands (isolated or non-isolated) are typically either fully filled or completely empty, lacking the conditions for spontaneous spin splitting and do not possess intrinsic magnetism [13, 17].

In this work, a dodecagonal cluster composed of 12 pentagons is confirmed to have two compact localized states (CLSs). Based on this large-size CLS-cluster, we designed a series of 3-coordinated two-dimensional (2D) lattices using the $RG^2$ method [36-38], and found that they all possess two corresponding flat-bands under the first-neighbor approximation of tight-binding model (TB) [39-41]. Interestingly, when considering these lattices as 2D carbon, they all exhibit relatively good thermodynamic stability, and most maintain fairly flat-bands even under multiple-neighbor long-range interactions. Among these structures, three systems with isolated flat-bands at the Fermi-level and three with weakly overlapped flat-bands (partially occupied) at the Fermi-level caught our significant interest. We further confirmed that two of them (191-8-66-C-r567x-1 and 191-10-90-C-r567x-1) are dynamically stable graphene allotropes and employed high-level HSE functionals (HSE06) to study their electronic and magnetic properties. The results indicate that 191-8-66-C-r567x-1 is a flat-band related magnetic half-metal and 191-10-90-C-r567x-1 is a flat-band related magnetic metal.

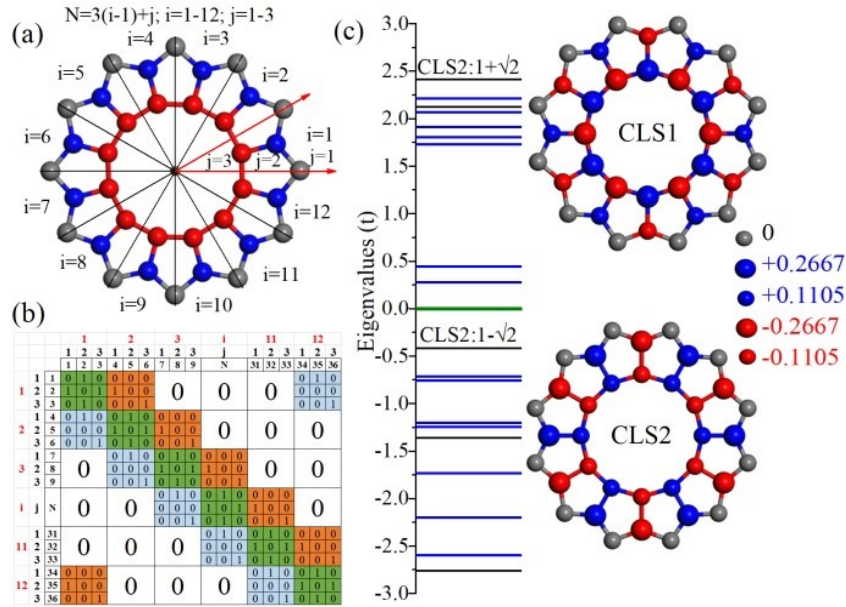

Fig. 1 (a) The dodecagonal cluster composed of 12 pentagons, which contains 3 inequivalent sites as marked in small gray, blue and red balls, respectively. (b) The schematic diagram of the 36-dimensional adjacency matrix for this cluster. (c) The eigenvalues of this 36-dimensional matrix, where the singlet, doublet, and triplet states are depicted by black, blue, and green lines, respectively. Additionally, it shows the projections of the eigenstates corresponding to the eigenvalues $1+\sqrt{2}$ and $1-\sqrt{2}$ to show their CLS distribution features.

## 2. Eigenvalues and eigenvectors of the CLS cluster

Fig. 1 (a) shows the dodecagonal cluster composed of 12 pentagons. It contains a total of 36 vertices distributed across three concentric regular dodecagons. This large dodecagon can be viewed in polar coordinates as a one-dimensional strip with a periodicity of $\theta=\pi/6$, containing 3 inequivalent positions within each period as denoted in gray (3.488, 0), blue (2.920, $\pi/12$) and red (1.949, $\pi/12$) balls. These coordinates ensure that each pentagon is nearly a regular pentagon with a side length of a=1. As shown in Fig. 1 (b), the 36 vertices are numbered as N=3(i-1)+j, in the usual polar coordinate sequence, where i and j are the periodicity index and the sub-index of inequivalent positions, respectively. We can construct the adjacency matrix of this dodecagonal cluster by using the side length (a=1) as the cutoff and it is a 36-dimensional sparse matrix, which can be represented using only three inequivalent 3×3 matrices:

$$A = \begin{vmatrix} 0 & 1 & 0 \\ 1 & 0 & 1 \\ 0 & 1 & 0 \end{vmatrix}, B = \begin{vmatrix} 0 & 0 & 0 \\ 1 & 0 & 0 \\ 0 & 0 & 1 \end{vmatrix}, O = \begin{vmatrix} 0 & 0 & 0 \\ 0 & 0 & 0 \\ 0 & 0 & 0 \end{vmatrix} \text{ and } C = B^\dagger = \begin{vmatrix} 0 & 1 & 0 \\ 0 & 0 & 0 \\ 0 & 0 & 1 \end{vmatrix}$$

And such a sparse matrix can be further decomposed into a 12-dimensional block matrix:

$$M = \begin{vmatrix} A & B & O & O & O & O & O & O & O & O & O & C \\ C & A & B & O & O & O & O & O & O & O & O & O \\ O & C & A & B & O & O & O & O & O & O & O & O \\ O & O & C & A & B & O & O & O & O & O & O & O \\ O & O & O & C & A & B & O & O & O & O & O & O \\ O & O & O & O & C & A & B & O & O & O & O & O \\ O & O & O & O & O & C & A & B & O & O & O & O \\ O & O & O & O & O & O & C & A & B & O & O & O \\ O & O & O & O & O & O & O & C & A & B & O & O \\ O & O & O & O & O & O & O & O & C & A & B & O \\ O & O & O & O & O & O & O & O & O & C & A & B \\ B & O & O & O & O & O & O & O & O & O & C & A \end{vmatrix}$$

We then computed and analyzed the eigenvalues and eigenvectors of this 36-dimensional matrix as the results shown in Fig. 1(c). There are 1 triplet (zero energy), 14 doublet and 5 singlet states as denoted in green, blue and red lines, respectively. We project these eigenstates into real space in order of increasing eigenvalue, as shown in Fig. S1. It is interesting that there are two eigenstates (1+√2 and 1-√2) that have zero projection onto all the outermost vertices (marked as gray balls in Fig. 1) and their projections onto the next layer alternate in sign with equal magnitudes. This alternating pattern satisfies the requirements for coherent interference and cancellation, indicating that such a dodecagonal cluster can be considered as a CLS cluster for designing 2D flat-band lattices.

### 3. Geometrical configurations and TB band structures of the lattices

As illustrated in Fig. S2 (a) and (b), we consider the two high-symmetry hexagonal arrangements of the CLS clusters and utilize RG$^2$ [36-38] to fill the vacuum regions between the clusters. To maintain similar geometrical features as those nodes in the CLS cluster, all newly introduced nodes must be three-coordinated. And RG$^2$ will ensure these nodes have similar distances and bond angles to achieve a uniform spatial distribution. Except for the large dodecagon at the center of the CLS-cluster, the entire lattice is only allowed to contain pentagons, hexagons, heptagons, and octagons. These geometric requirements are essential for enabling the subsequent development of the lattice into 2D carbon allotropes. Finally, we discover 38 non-equivalent 2D lattices that meet these criteria, as their configurations are shown in Fig. S4. The TB model with the nearest-neighbor approximation is employed to study the dispersion relations of these lattices and only $p_z$ orbitals are considered in our calculations. The Hamiltonian can be written as $H = -\sum_{i \neq j} t_{ij}(c_i^\dagger c_j + hc) + \varepsilon_\pi \sum_i c_i^\dagger c_i$, where $t_{ij}$ is the hopping integral and $\varepsilon_\pi$ is the onsite energy, $c_i^\dagger$ and $c_j$ denote the creation and annihilation operators, respectively. For lattice models, the overlap terms

($S = \sum_{i \neq j} S_{ij}(c_i^\dagger c_j + hc) + \sum_i c_i^\dagger c_i$) are ignored and the onsite energies $\varepsilon_\pi$ are set to be zero. All the nearest-neighbor integrals $t_{ij}$ are set to be -1 eV and other terms are all zero. The band structures of all these 38 lattices can be quickly calculated by our gt-TB code [39-41] as shown in Fig. S5. As we anticipated, both the CLS states are well-preserved in these 2D periodic lattices. They retain their original eigenvalues (1+√2 and 1-√2) and appear as flat-bands (red solid lines) in the energy band structures of these 2D lattice models.

## 4. Crystal structures and TB band structures of the graphene allotropes

To realize these 2D lattice models in real chemical systems, we naturally consider sp$^2$ carbon due to its rich diversity of allotropes with three-coordinated configurations. As we know, carbon can form various crystal structures with rich electronic properties, even in 2D sp$^2$ allotropes, including semiconductors [40, 42], metals [40, 43], Dirac cones [38, 40, 44], and Dirac rings [40]. However, the reports on flat-band and magnetism in 2D carbon allotropes remain scarce [17, 29]. Here, we investigate whether these potential 2D carbon allotropes can maintain flat-bands under multiple neighbor interactions, particularly focusing on the isolated zero-energy flat-bands near the Fermi-level. We adjust the distances between lattice points to match the C-C bond length in graphene and use first-principles methods based on density functional theory (DFT) [45, 46] to further optimize these 2D graphene allotropes. In our calculations, GGA-PBE [47] pseudo-potential is employed for structure optimization and HSE06 [48] functional is used to confirm the electronic properties. The cutoff energy of the wave function basis is set to be 500 eV and the k-point sampling grid is designed based on the lattice constant (0.211 /Å). Both the lattice constants and atomic positions are rigorously optimized until the force on each atom is less than 0.02 eV/Å.

All the optimized crystal structures of these 2D carbon allotropes and their corresponding average energies relative to graphene are shown in Fig. S4. Following our earlier work [36, 39, 40], we designate these structures based on symmetry and topological rings as sn-in-tn-C-rs-id. Here, sn represents the space group number, in and tn indicate the number of inequivalent atoms and total atoms, C denotes the chemical formula, rs stands for the ring series contained in the system, and an id number associated with the energy order ensures the uniqueness of the structure name. Except for one in space group No.187, they are predominantly distributed in space groups of No. 175 and No.191. As shown in Fig. S3, most of these 2D carbons are energetically more stable the experimentally synthesized Phagraphene [44], TPH-graphene [49] and Biphenylene [50, 51]. These results suggest a high probability of synthesizing our newly discovered 2D carbon allotropes in future experiments.

To investigate the electronic band structures of these potential graphene allotropes, the overlap terms $S_{ij}$ and the long-range hopping terms are included in our calculations based on gt-TB code [39-41]. The distance-dependent hopping and overlap are described as $t_{ij} = t_0 e^{q_1(1-\frac{d_{ij}}{d_0})}$ and $S_{ij} = s_0 e^{q_2(1-\frac{d_{ij}}{d_0})}$, where $d_0$ represents the standard bond length in graphene, which is 1.426 Å. The variable $d_{ij}$ denotes the distance between the $i$th and $j$th atoms in the system. The parameters $t_0$ and $s_0$ correspond to the hopping and overlap integrals at ideal distance $d_0$, respectively. These integrals will decay exponentially as the distance $d_{ij}$ increase, with the decay rates governed by $q_1$ and $q_2$. For distance $d_{ij}$ greater than $d_{cut} = 10$ Å, all $t_{ij}$ and $S_{ij}$ are considered to be zero. Additionally, the onsite energies $\varepsilon_\pi$ are set to be zero for all carbon p$_z$ orbitals in this work. Usually, the PBE or HSE based band structures are needed for fitting these TB parameters. Our previous work [39, 40] has already successfully completed the universal parameter fitting for the HSE06 based band structures of 2D sp$^2$ carbons with highly effective results. Here, the HSE-based parameters ($t_0 = -3.02\ eV, s_0 = 0.235\ eV, q_1 = 2.11\ and\ q_2 = 2.01$) are directly used for a rapid band structure calculations for these newly discovered 2D carbon structures.

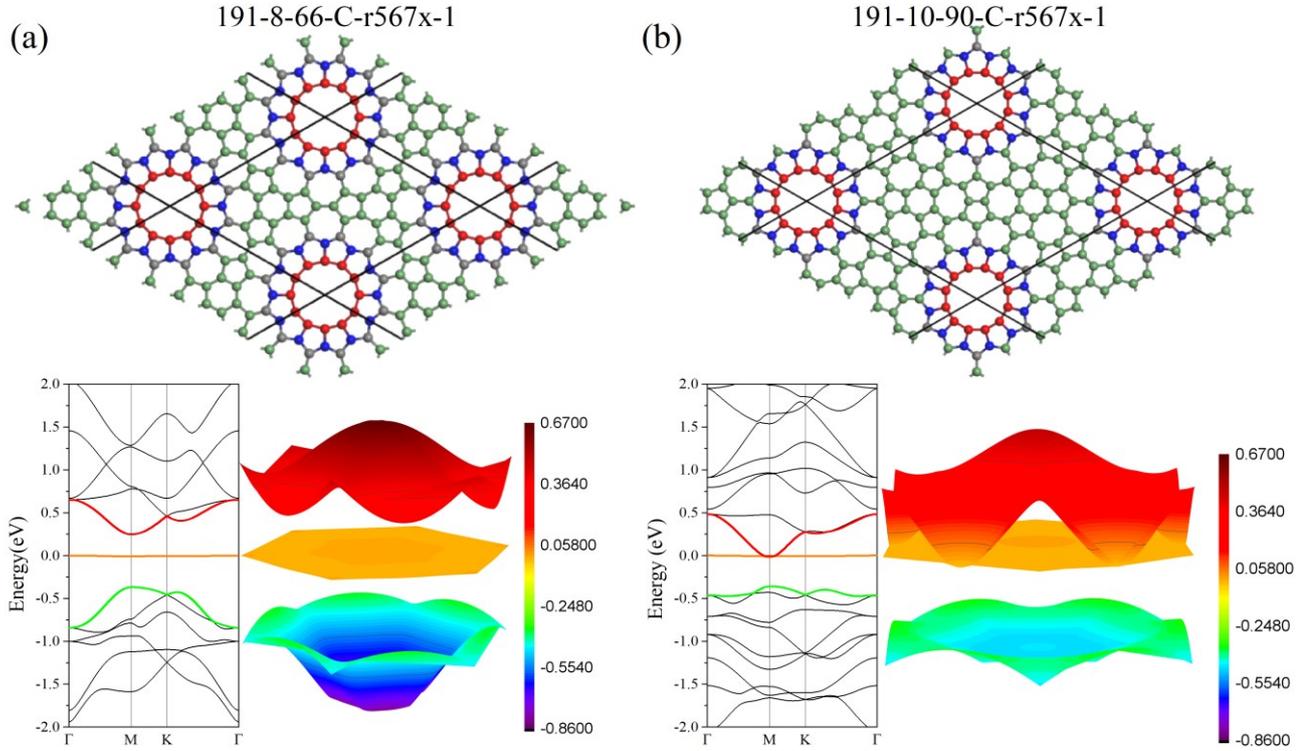

Fig. 2 The optimized crystal structures of 191-8-66-C-r567x-1 (a) and 191-10-90-C-r567x-1 (b). Their TB band structures calculated from HSE06-based parameters are also provided in 2D and 3D views.

As shown in Fig. S6, most of these graphene allotropes exhibit flat-bands at different energy positions in their band structures, though the flatness of these bands is variably influenced by long-range interactions. And it is easy to notice that most of the flat-bands are located nearby the Fermi-level. Notably, as revealed in Fig.S7, there are at least three of these systems (191-8-66-C-r567x-1, 191-9-84-C-r567x-1 and 175-17-102-C-r567x-1) possess isolated flat-bands at the Fermi-level and three others (191-10-90-C-r567x-1, 175-17-102-C-r5678x-3 and 175-15-86-C-r567x-1) exhibit weakly overlapped flat-bands at the Fermi-level. However, the simulated vibrational spectrums as shown in Fig.S8 suggest that only 191-8-66-C-r567x-1 and 191-10-90-C-r567x-1 are dynamically stable as 2D carbon allotropes. Thus, we further calculated the electronic band structures of 191-8-66-C-r567x-1 and 191-10-90-C-r567x-1 in different functionals of normal PBE (Fig.S9) and high-level HSE06 (Fig.3). Both PBE and HSE06 show that 191-8-66-C-r567x-1 is a flat-band insulator and 191-10-90-C-r567x-1 is flat-band metal. These results are consistent with the TB results as shown in Fig.2, in which the 3D band structures suggested that the flat-bands in these two systems are flat throughout the entire BZ. The Fermi velocities for the flat-band states in 191-8-66-C-r567x-1 and 191-10-90-C-r567x-1 are calculated to be $1\times10^4$ m/s and $0.786\times10^4$ m/s, respectively, which are smaller than those in magic-angle graphene [5] of $4\times10^4$ m/s. These results indicate that the flatnesses of these flat-bands are close to those in magic-angle graphene bilayers. Finally, as the projected charge densities of the flat-bands shown in Fig.S10, it is clear that both the flat-bands in 191-8-66-C-r567x-1 and 191-10-90-C-r567x-1 are obviously localized in the dodecagonal clusters as CLS states.

## 5. The magnetic and electronic properties of 191-8-66-C-r567x-1 and 191-10-90-C-r567x-1

It is believed that isolated and weakly overlapped flat-bands at the Fermi-level will introduce spin-splitting in the systems [13, 17, 34]. To confirm whether 191-8-66-C-r567x-1 and 191-10-90-C-r567x-1 are intrinsic magnetic phases, their spin-splitting band structures are calculated in different functionals of normal PBE and high-level HSE06 as shown in Fig.S8 and Fig.3, respectively. There are obvious spin-splitting phenomena that can be noticed in the band structures. It is easy to see that PBE results suggest that both 191-8-66-C-r567x-1 and 191-10-90-C-r567x-1 are flat-band related magnetic metals, while HSE06 shows that 191-8-66-C-r567x-1 is a magnetic half-metal and 191-10-90-C-r567x-1 is a normal magnetic metal. As the results summarized in Table S1, we can see that both PBE and HSE suggest that ferromagnetic states are always more energetically stable than the nonferromagnetic states for both 191-8-66-C-r567x-1 and 191-10-90-C-r567x-1. For example, HSE suggests that the ferromagnetic state in 191-10-90-C-r567x-1 is about 12 meV/magnet lower than the nonferromagnetic state. Although the average energy difference of each atom is just a few meV, we still believe that magnetic carbon phases have been designed in our present work based on the flat-band (CLS) theory. In PBE calculation, the total magnetic moments of 191-8-66-C-r567x-1 and 191-10-90-C-r567x-1 are calculated to be 0.784 μB/cell and 0.522 μB/cell, respectively. And these values will be confirmed to be 1.854 μB/cell and 1.663 μB/cell in high-level HSE06 based calculations.

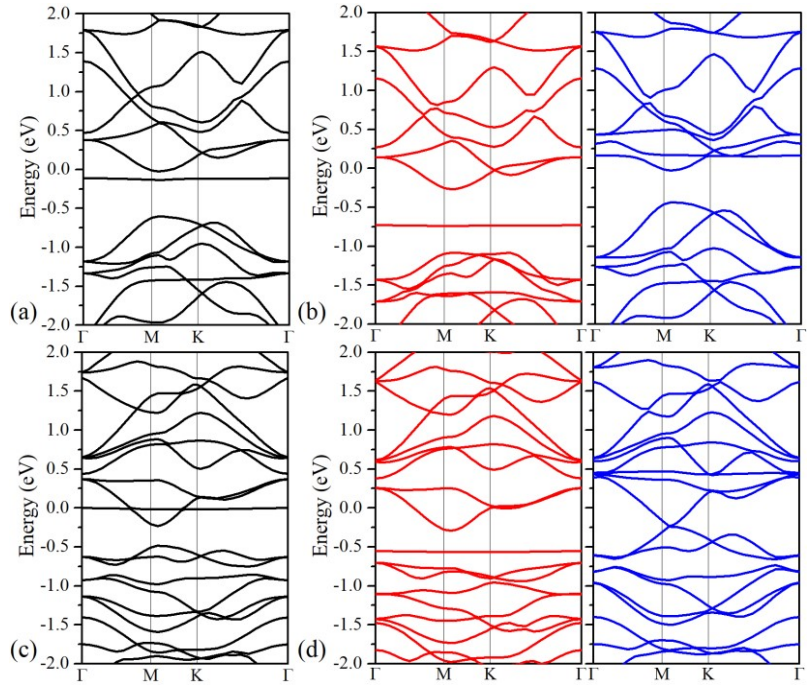

Fig. 3. HSE-based band structures for 191-8-66-C-r567x-1 and 191-10-90-C-r567x-1in nonferromagnetic (black solid lines) and ferromagnetic (red and blue solid lines) states.

## 6. Conclusion

In a word, a special dodecagonal cluster composed of 12 pentagons has been confirmed to be a good CLS cluster, based on which a total of 38 three-coordinated double flat-band lattices have been discovered for designing two-dimensional flat-band materials. As two-dimensional carbon crystals, most of these lattices maintain good flat-band characteristics even after multiple nearest-neighbor interactions. Especially, in some systems, the flat-bands appear isolated at the Fermi-level or only have weak overlap with other dispersive bands. Two of them, named as 191-8-66-C-r567x-1 and 191-10-90-C-r567x-1, have been confirmed to be dynamically stable carbon phases by first-principles calculations. The maximum Fermi velocities of the flat-band electrons in 191-8-66-C-r567x-1 ($1\times10^4$ m/s) and 191-10-90-C-r567x-1 ($0.786\times10^4$ m/s) are lower than those in magic-angle graphene ($4\times10^4$ m/s), which indicate that flat-bands in these carbon monolayers are flatter than those in magic-angle graphene. Finally, we have confirmed that 191-8-66-C-r567x-1 is a flat-band related magnetic half-metal, while 191-10-90-C-r567x-1 is a flat-band related magnetic normal metal. Our present work shows that flat-bands can be realized in carbon monolayer and this provides a potential way to realize metal-free magnetism in light element systems.

## Acknowledgments

This work is supported by the National Natural Science Foundation of China (Grant Nos. 52372260, and 12204397, 12374046), the Youth Science and Technology Talent Project of Hunan Province (Grant No. 2022RC1197) and the Research Foundation of Education Bureau of Hunan Province, China (Grant No. 23A0102) and the Science Fund for Distinguished Young Scholars of Hunan Province of China (No.2024JJ2048).

# Supplementary for "Isolated zero-energy flat-bands and intrinsic magnetism in carbon monolayers"

Chaoyu He,[1,2,*] Shifang Li,[1] Yuwen Zhang[1], Zhentao Fu[2], Jin Li[1] and Jianxin Zhong[2,†]

[1] School of Physics and Optoelectronics, Xiangtan University, Xiangtan 411105, China

[2] Center for Quantum Science and Technology, Shanghai University, Shanghai 200444, China

Email of Corresponding Author: hechaoyu@xtu.edu.cn; jxzhong@xtu.edu.cn

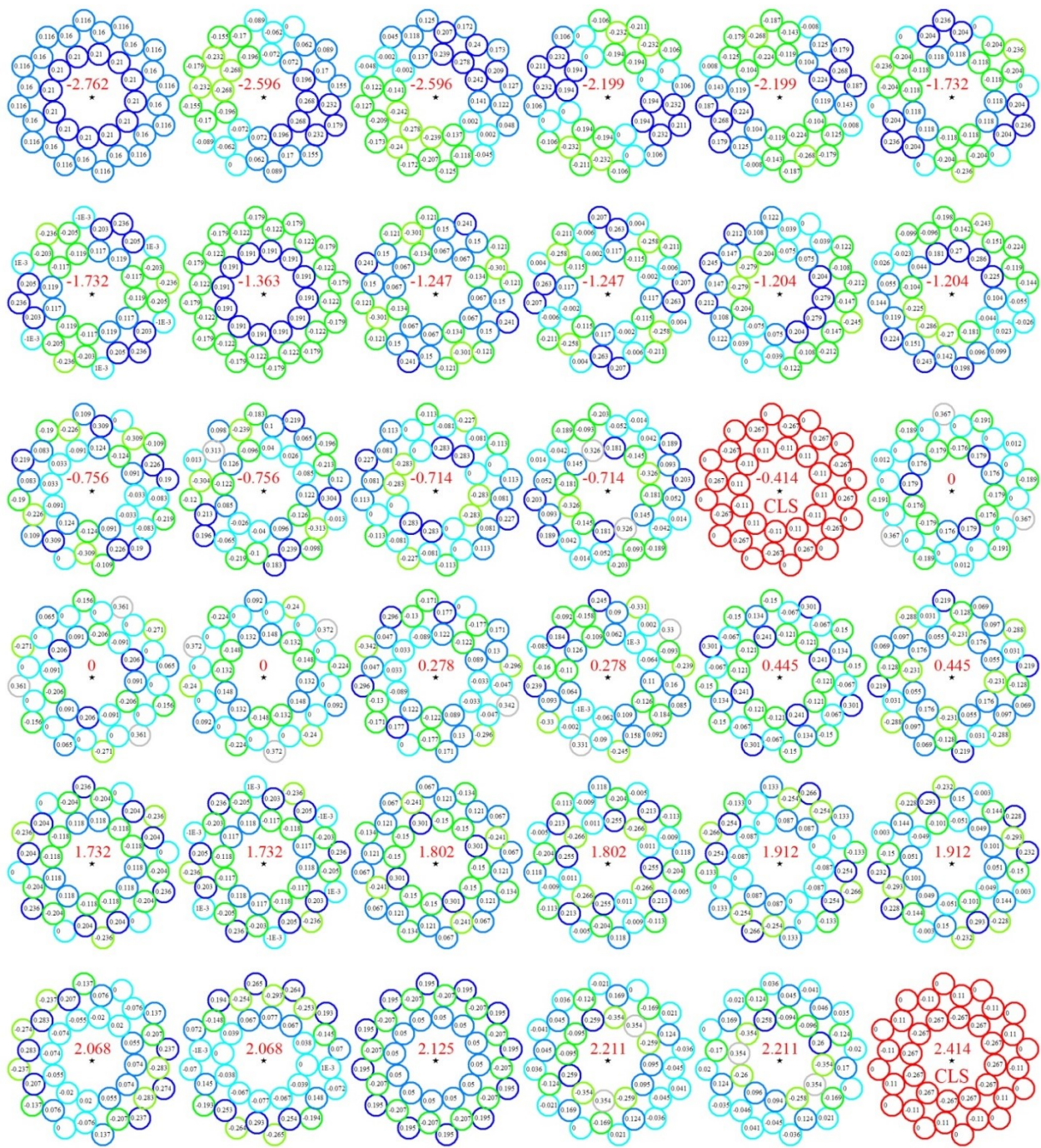

Fig. S1 The projections of all the 36 eigenstates of the dodecagonal cluster. The two CLS states are marked as pure red circles, while circles for other states are colored according to their contributions to the corresponding eigenstates.

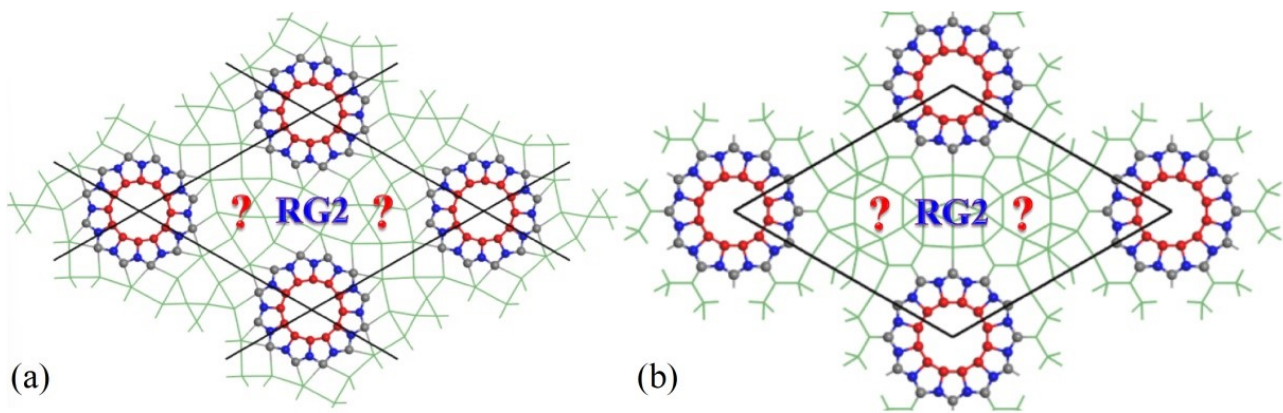

Fig. S2. The two high-symmetry arrangements of the dodecagonal clusters for designing potential 2D flat-band lattices.

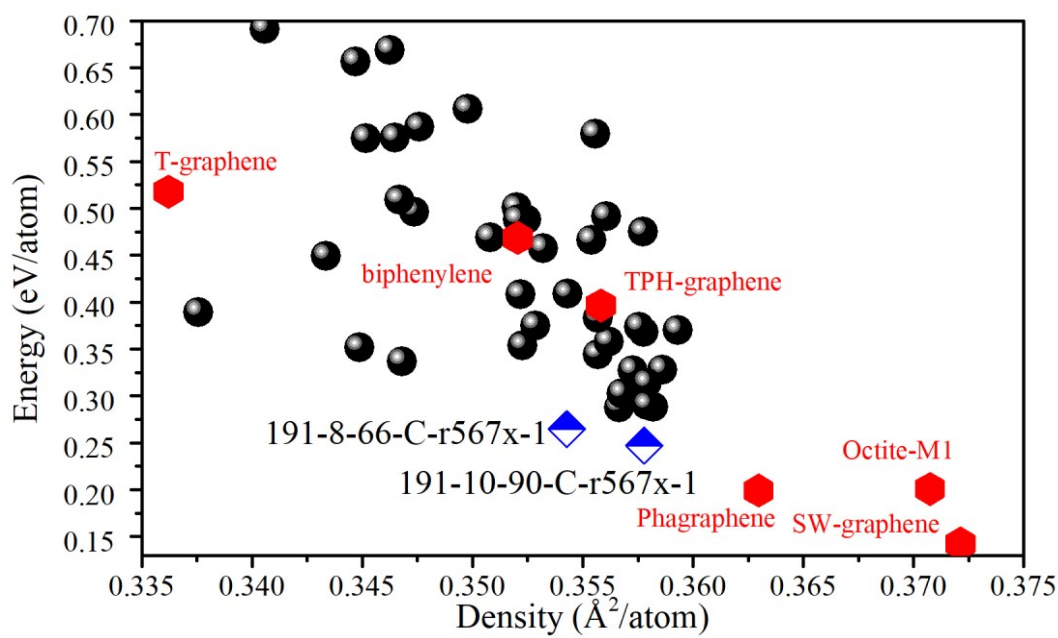

Fig. S3. The average energies against carbon densities of these 2D carbon allotropes (Black solid balls and blue solid squares), where the energy of graphene is set to zero as a reference and some previously reported graphene allotropes are considered for comparison (Red sold hexagons).

Table 1 The calculated total energies, magnetic moments and electronic properties of 191-8-66-C-r567x-1 and 191-10-90-C-r567-1 based on different functionals and magnetic settings.

|  | 191-8-66-C-r567x-1 | | 191-10-90-C-r567x-1 | |
|---|---|---|---|---|
|  | Energy | Electronic property | Energy | Electronic property |
| HSE-TB-NM | - | Flat-band insulator |  | Flat-band metal |
|  |  |  |  |  |
| HSE-NM | -728.448 eV/cell<br>-11.0731 eV/atom<br>-60.704 eV/magnet | Flat-band insulator | -994.833 eV/cell<br>-11.0533 eV/atom<br>-82.902 eV/ magnet | Flat-band metal |
| HSE-FM | -728.464 eV/cell<br>-11.0737 eV/atom<br>-60.705 eV/magnet | Flat-band half-metal<br>1.854 μB/cell | -995.023 eV/cell<br>-11.0558 eV/atom<br>-82.918 eV/magnet | Flat-band metal<br>1.663 μB/cell |
|  |  |  |  |  |
| PBE-NM | -591.223 eV/cell<br>-8.9579 eV/atom<br>-49.268 eV/magnet | Flat-band metal | -807.756 eV/cell<br>-8.9750 eV/atom<br>-67.313 eV/magnet | Flat-band metal |
| PBE-FM | -591.229 eV/cell<br>-8.9580 eV/atom<br>-49.269 eV/magnet | Flat-band metal<br>0.784 μB/cell | -807.761 eV/cell<br>-8.9751 eV/atom<br>-67.314 eV/magnet | Flat-band metal<br>0.522 μB/cell |

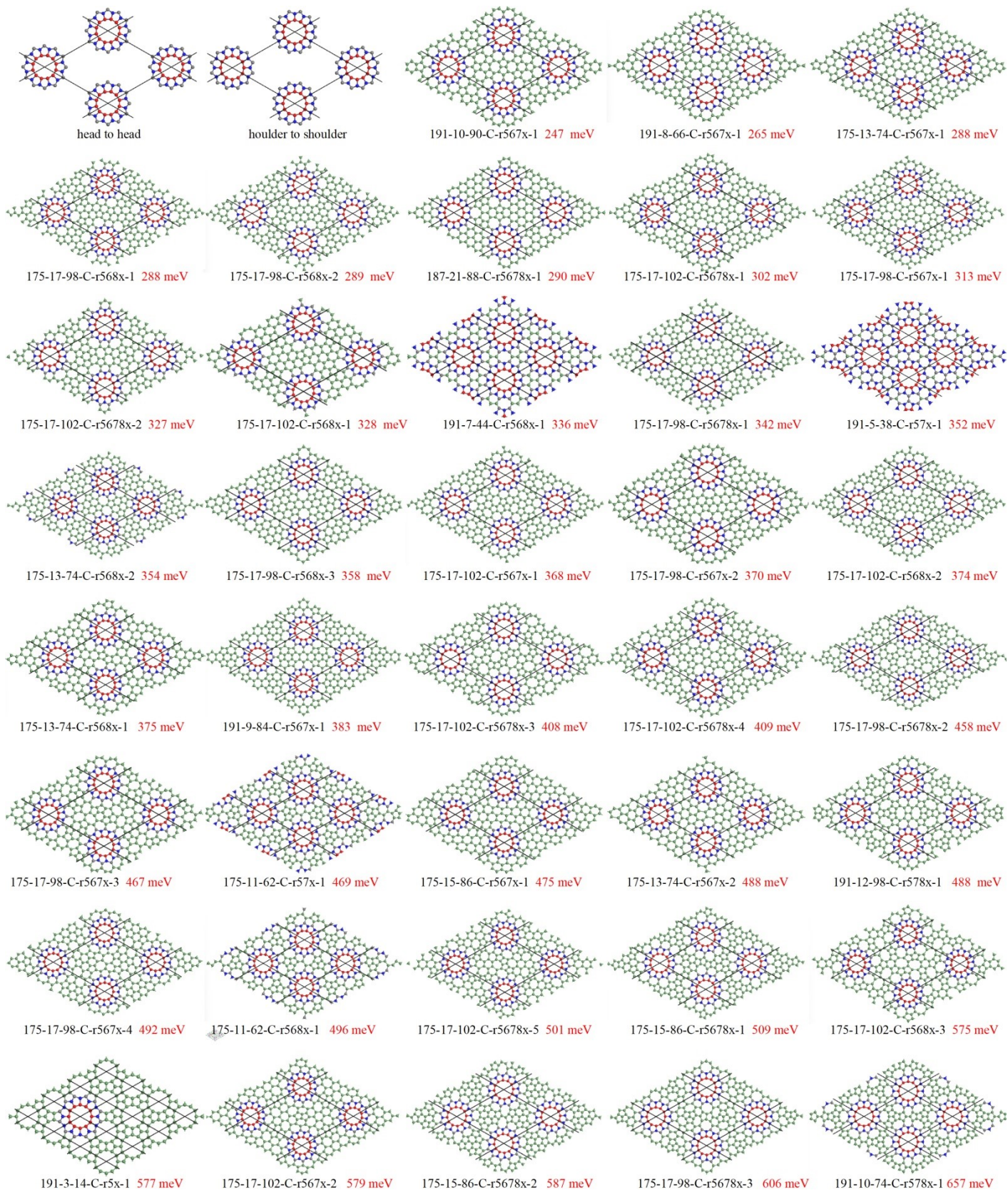

Fig. S4. The two different stacking manners of the dodecagonal clusters in hexagonal symmetry and the 38 structure candidates with only 3-coordinated notes for connecting the dodecagonal cluster to 2D lattices. All these configurations are fully optimized as graphene allotropes. Their names and average energies per atom relative to graphene are included below their corresponding crystal structures.

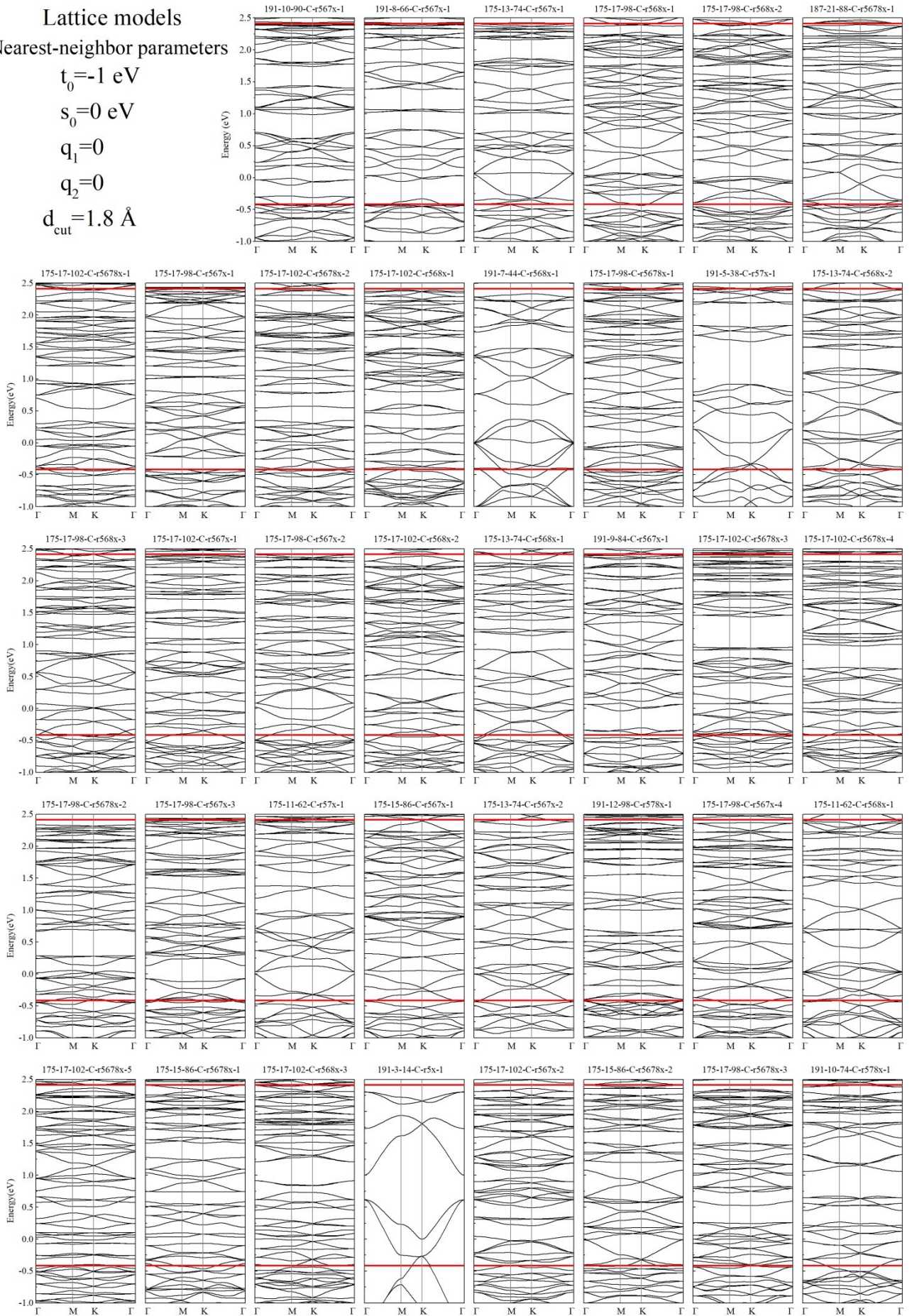

Fig. S5. The band structures for all the 38 candidate lattices calculated by gt-TB code with considering only the first-neighbor interaction (t=-1 eV).

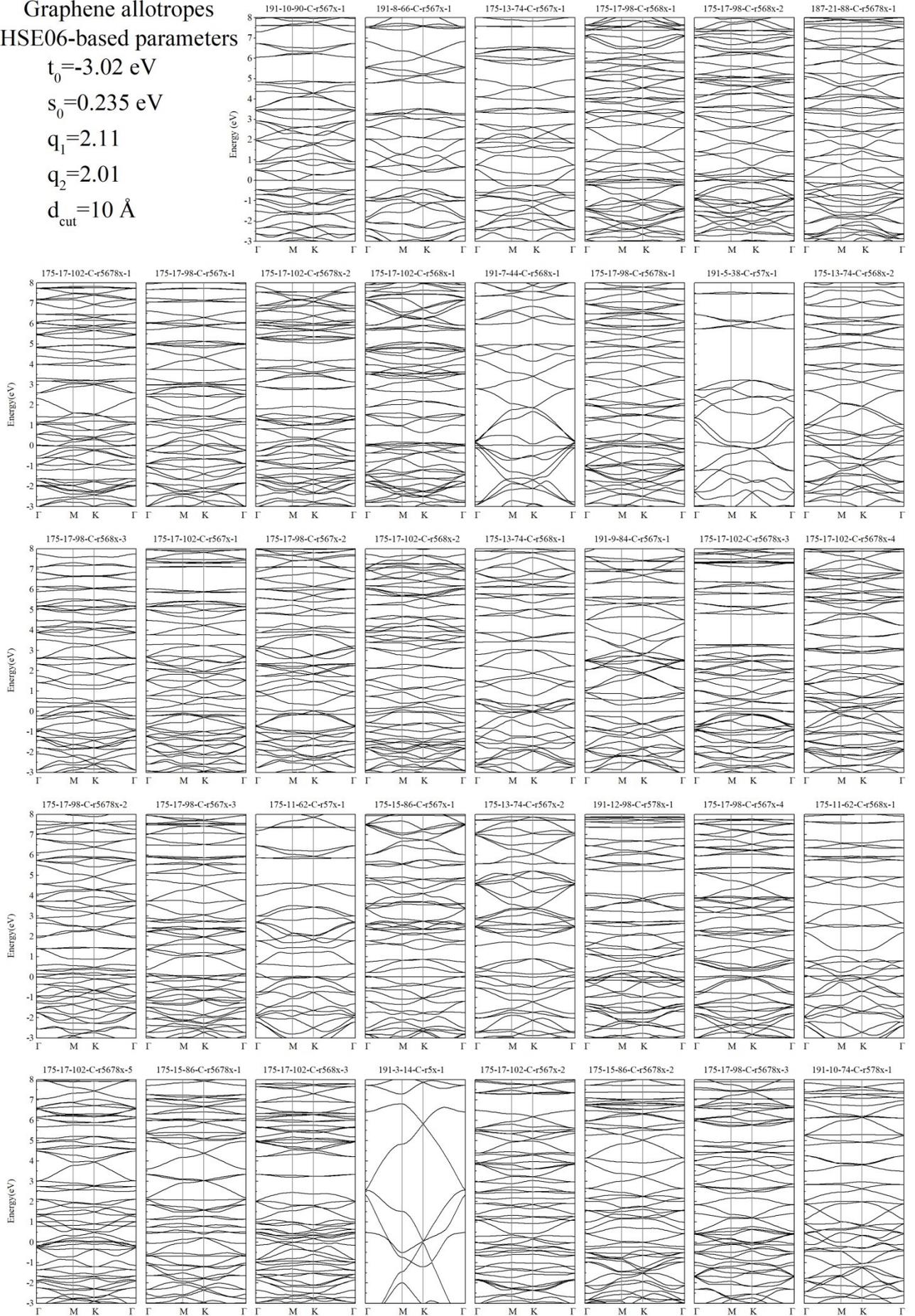

Fig. S6. The band structures for all the 38 graphene allotropes calculated by gt-TB with considering all interactions within 10 Å. The corresponding HSE-based parameters are set to $t_0=-3.02$ eV, $s_0=0.235$ eV, $q_1=2.11$ and $q_2=2.01$.

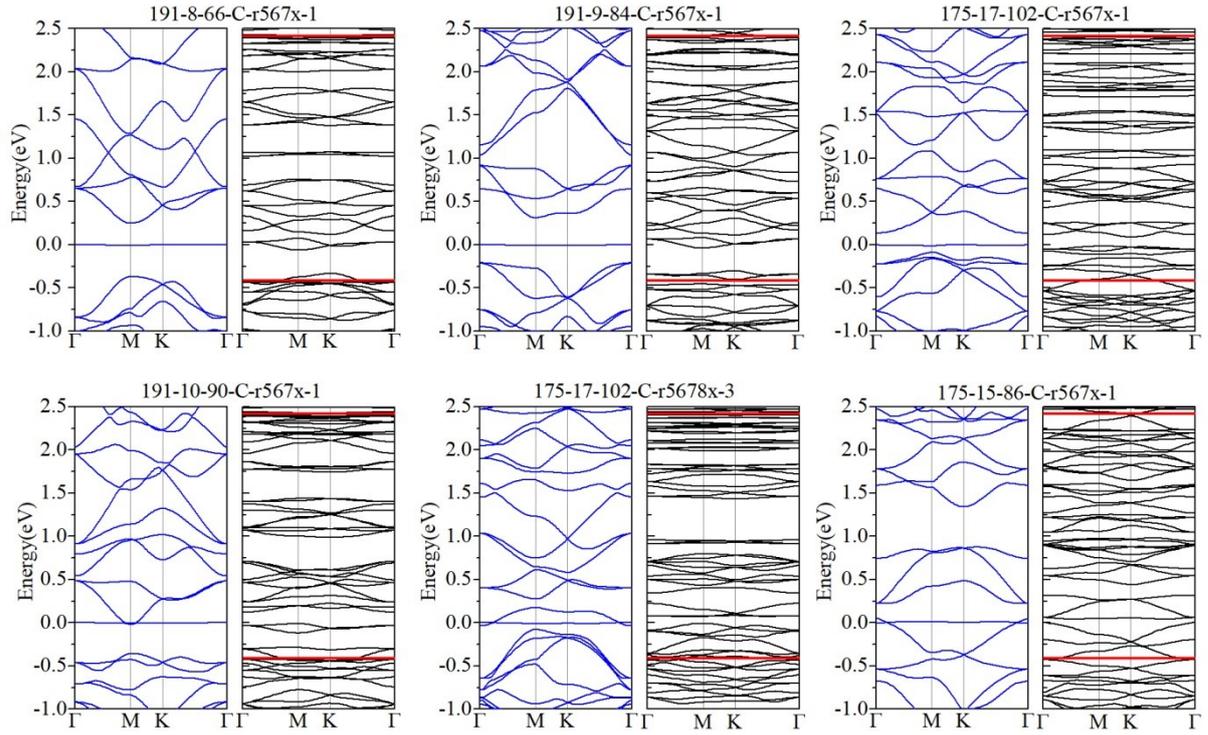

Fig. S7. The comparison in band structures calculated based on only first-neighbor interactions (black solid lines) and certain long-range interactions (blue solid lines) for those graphene allotropes with isolated or weakly overlapped flat-bands at the Fermi-level.

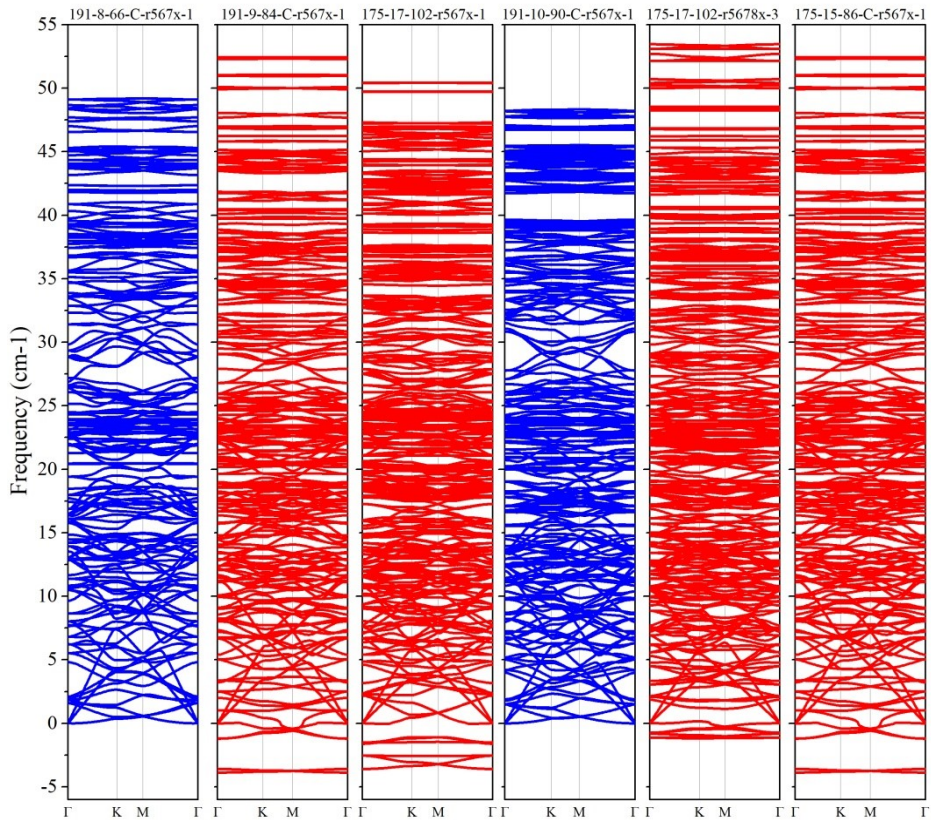

Fig. S8. PBE-based phonon band structures for the six graphene allotropes with isolated or weakly overlapped flat-bands at the Fermi-level.

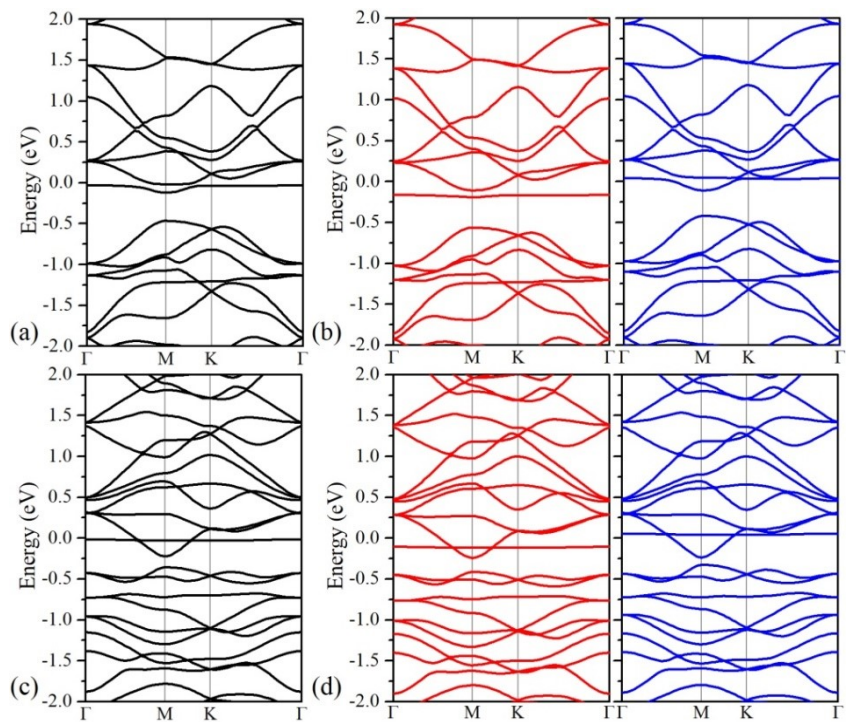

Fig. S9. PBE-based band structures for 191-8-66-C-r567x-1 and 191-10-90-C-r567x-1 in nonferromagnetic (black solid lines) and ferromagnetic (red and blue solid lines) states.

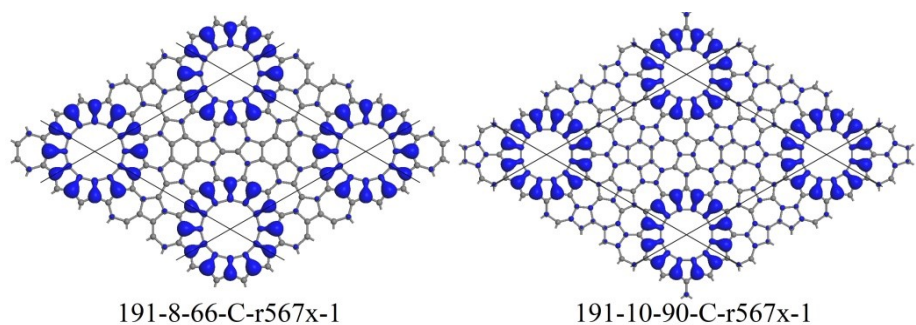

191-8-66-C-r567x-1  191-10-90-C-r567x-1

Fig. S10. PBE-based results: The projected charge density for the flat-band near the Fermi-level (blue solid surface) in 191-8-66-C-r567x-1 and 191-10-90-C-r567x-1.

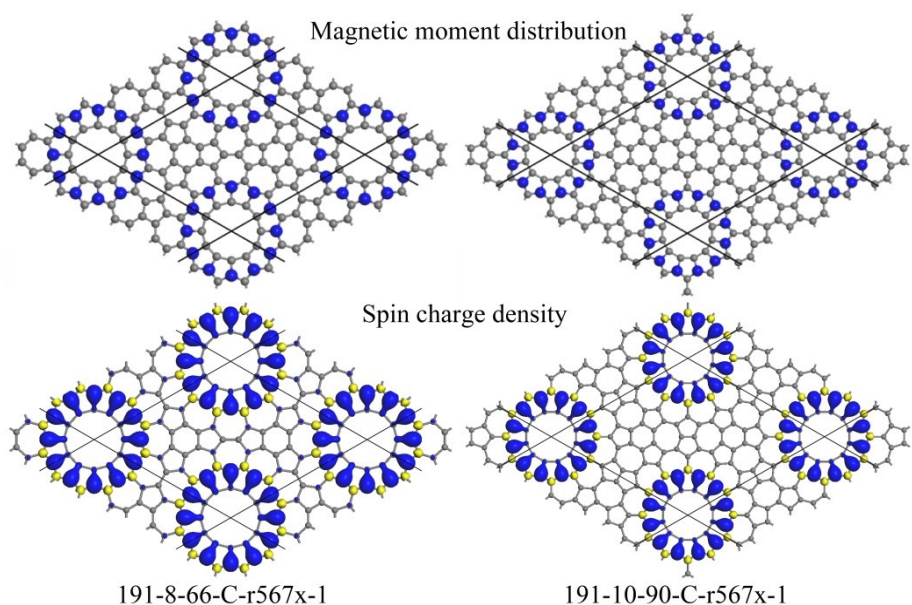

191-8-66-C-r567x-1  191-10-90-C-r567x-1

Fig. S11. PBE-based results: The distributions of magnetic moments (blue solid balls) and spin charge density (blue surface for spin up and yellow surface for spin down) in 191-8-66-C-r567x-1 and 191-10-90-C-r567x-1.